\documentclass[%
reprint,
 amsmath,amssymb,
 aps,prl]{revtex4-1}

\usepackage{float}
\usepackage{graphicx}
\usepackage{dcolumn}
\usepackage{bm}
\usepackage[flushleft]{threeparttable}
\usepackage{url}
\usepackage{amsmath}
\usepackage{amssymb}
\usepackage{braket}
\usepackage{multirow}
\usepackage{array}
\usepackage{url}
\newcolumntype{P}[1]{>{\centering\arraybackslash}p{#1}}
\usepackage{xcolor}
\usepackage{framed}
\usepackage{lipsum}
\colorlet{shadecolor}{blue!60}
\usepackage{soul}
\usepackage{hyperref}
\usepackage{upgreek}
\usepackage{ulem}
\makeatletter
\newcommand*{\rom}[1]{\expandafter\@slowromancap\romannumeral #1@}
\makeatother

\begin{document}

\preprint{APS/123-QED}

\title{Stabilization of competing ferroelectric phases of HfO$_2$ under epitaxial strain}

\author{Yubo Qi$^1$, Sobhit Singh$^1$, Claudia Lau$^2$, Fei--Ting Huang$^1$, Xianghan Xu$^1$ \\
Frederick J. Walker$^2$, Charles H. Ahn$^2$, Sang--Wook Cheong$^1$, and Karin M. Rabe$^1$}

\affiliation{%
$^1$Department of Physics $\&$ Astronomy, Rutgers University, \\
Piscataway, New Jersey 08854, USA \\
$^2$Department of Physics, Yale University,\\ New Haven, Connecticut 06520, USA
}%

\date{\today}

\begin{abstract}

Hafnia (HfO$_2$)--based thin films have promising applications in nanoscale electronic devices due to their robust ferroelectricity and integration with silicon. 
Identifying and stabilizing the ferroelectric phases of HfO$_2$ have attracted intensive research interest in recent years. 
In this work, first--principles calculations on (111)--oriented HfO$_2$ are used to 
discover that imposing an in--plane shear strain on the metastable tetragonal phase drives it to a polar phase. 
This in--plane--shear--induced polar phase is shown to be an epitaxial--strain--induced distortion of a previously--proposed metastable ferroelectric $Pnm2_1$ phase of HfO$_2$. 
This ferroelectric $Pnm2_1$ phase can account for the recently observed ferroelectricity in (111)--oriented HfO$_2$--based thin films on a SrTiO$_3$ (STO) (001) substrate [\href{https://www.nature.com/articles/s41563-018-0196-0}{Nature Materials 17, 1095--1100 (2018)}].
Further investigation of this alternative ferroelectric phase of HfO$_2$ could potentially improve the performances of HfO$_2$--based films in logic and memory devices.

\end{abstract}

\pacs{Valid PACS appear here}

\maketitle


Ferroelectrics are materials with spontaneous electric polarization that can be switched by the application of an external electric field.
This property makes ferroelectrics useful for a wide range of practical applications, such as non--volatile memory devices~\cite{Kingon00p1032}, 
field effect transistors~\cite{Mathews97p238}, and tunable capacitors~\cite{Padmini99p3186,Pervez04p4451}.
However, most conventional ferroelectrics, such as BaTiO$_3$ or PbTiO$_3$, are not suitable for nanoscale devices, 
because of the depolarization field effect, which suppresses the ferroelectricity and becomes more significant as the thickness of films decreases~\cite{Dawber03p393,Wurfel73p5126,Kim05p237602}.
HfO$_2$--based materials are exceptions; the Al~\cite{Mueller12p2412}, Gd~\cite{Mueller12p123}, Sr~\cite{Pevsic16p4601}, Y~\cite{Muller11p114113,Olsen12p082905} doped HfO$_2$ and Hf$_{0.5}$Zr$_{0.5}$O$_2$ alloy~\cite{Muller11p112901} can sustain ferroelectricity in films thinner than 20 nm. Furthermore, HfO$_2$-based materials can be integrated with silicon processing, and indeed are currently used as gate dielectrics~\cite{ByoungLee_1999, Kingon00p1032, schaeffer2004hafnium}.

\begin{figure}[hpbt]
\includegraphics[width=7.0cm]{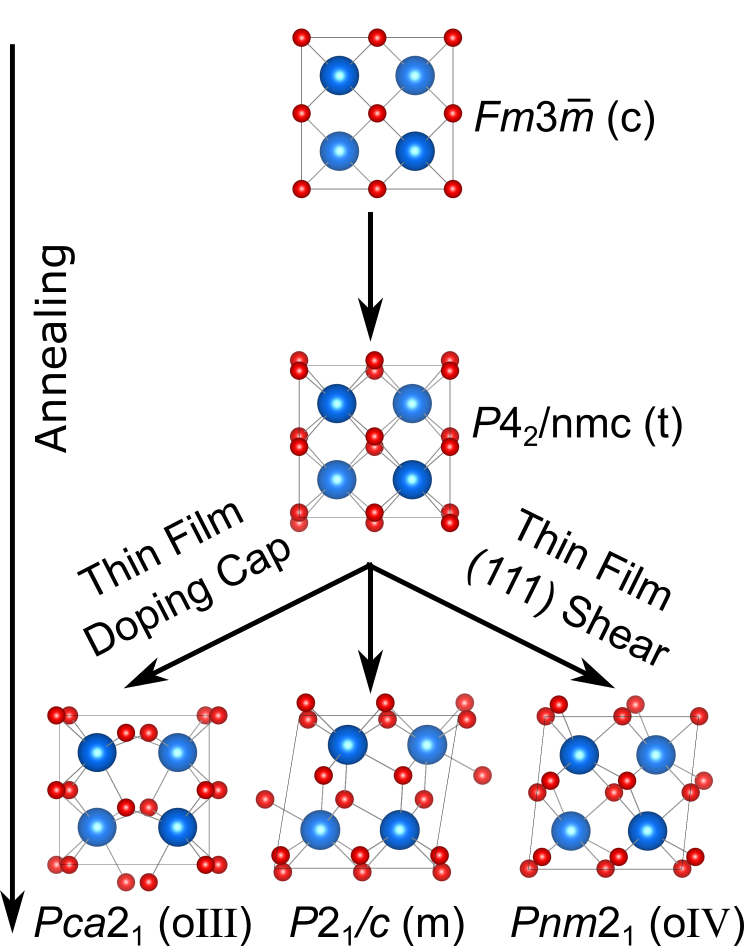}
\caption{Structures and transitions of the c, t, m, o\rom{3}, and o\rom{4} phases.
}\label{fig:f1}
\end{figure}

HfO$_2$ adopts various polymorphs~\cite{Huan14p064111,Pevsic16p4601} (Fig. \ref{fig:f1}). The ferroelectricity in thin films has generally been attributed to the formation of the orthorhombic $Pca2_1$ phase (o\rom{3}--phase) ~\cite{Boscke11p102903,Boscke11p112904,Sang15p162905}.
The formation of this phase has been shown to be affected by various extrinsic factors, 
such as pressure~\cite{Huan14p064111}, strain~\cite{Batra17p4139,Shiraishi16p262904}, dopants~\cite{Batra17p9102,Materlik18p164101,Schroeder14p08LE02,Starschich17p333,Park17p4677}, oxygen vacancies~\cite{Xu16p091501}, surface energies~\cite{Batra16p172902,Materlik15p134109,Park17p9973},  and electric fields~\cite{Batra17p4139,Reyes14p140103,Muller12p4318}. 
This attribution does not preclude the existence of other competing ferroelectric phases~\cite{Huan14p064111}.
For example, a ferroelectric phase different from o\rom{3} has been experimentally observed in (111)--oriented Hf$_{0.5}$Zr$_{0.5}$O$_2$ thin films~\cite{Wei18p1095}.
This (111)--oriented ferroelectric film has a distinctive field--cycling behavior; wake--up cyclings~\cite{Pevsic16p4601,Olsen12p082905} are not required for acquiring a steady $P$--$E$ hysteresis loop.
This intriguing observation invites a definitive identification of this novel phase, which should also provide a natural explanation for the lack of wake--up behavior and may be of great significance for practical applications of hafnia--based materials.

In this study, we carry out density--functional theory (DFT) based first-principles calculations on different polymorphs of HfO$_2$, with a particular focus on the ferroelectric orthorhombic $Pnm2_1$ phase (o\rom{4}--phase). 
Although this o\rom{4}--phase is energetically less favorable in bulk~\cite{Huan14p064111}, we provide evidence that it can be kinetically stabilized in epitaxial films with (111)--orientation via a transition from the tetragonal $P4_2/nmc$ phase (t--phase), which has compatible lattice parameters for (111)--oriented crystals. 
In fact, in the epitaxial geometry corresponding to a STO (001) substrate,
the nonpolar t--phase is no longer locally stable, and collapses into a distorted version of the ferroelectric o\rom{4} phase, with a robust ferroelectric polarization ($P=0.41$\,C/m$^2$).
Our simulated  x--ray diffraction (XRD) and selected area electron diffraction (SAED) for this phase are consistent with the ones for the polar phase of Hf$_{0.5}$Zr$_{0.5}$O$_2$ experimentally reported in Ref.~\cite{Wei18p1095}, leading to a greater understanding of functional behavior arising from competing ferroelectric phases in HfO$_2$--based thin films.

 We performed first-principles calculations of the energies, structural parameters and polarization of various phases of HfO$_2$ using the \textsc{Quantum--espresso}~\cite{Giannozzi09p395502etalp} plane--wave DFT code within the generalized gradient approximation (GGA). 
 The numerical details are provided in the Supplementary Materials (SM) section \rom{1}~\cite{SM}.
To understand the relationship between the tetragonal and the o\rom{4} phase,  we start with the bulk t--phase of HfO$_2$ oriented along the (111) direction, with in--plane lattice parameters $a=b$ = 7.24\,\AA, and cell angle $\gamma$ = 120.0\,$^\circ$. 
Then, keeping $a=b$ fixed, we vary $\gamma$ in range [120$^\circ$, 125$^\circ$], relaxing the structures to compute the potential energy profile as a function of the cell angle $\gamma$.

In Table~\ref{t1}, we list the computed pseudocubic lattice parameters and relative energies for several reported phases of HfO$_2$.
A schematic representation of the phase transitions among these phases is shown in Fig.~\ref{fig:f1}.
In bulk, HfO$_2$ adopts a highly--symmetric $Fm\overline{3}m$ cubic fluorite structure (c--phase) at high temperatures ($T>2773$ K),
and transforms to the t--phase as temperature decreases (2073 K$<T<$2773 K).
As the temperature drops below 2073 K, bulk HfO$_2$ transforms to a $P2_1/c$ monoclinic phase (m--phase)~\cite{Ruh68p23}.
Our DFT calculations show that among these three phases, the c--phase has the highest energy and the m--phase has the lowest energy (see Table~\ref{t1}), which is consistent with the experimental observations ~\cite{Ruh68p23} and previous calculations.

\begin{table}[hpbt]
\begin{center}
 \begin{tabular}{ P{2.0cm} P{0.8cm} P{0.8cm} P{0.8cm} P{0.8cm} P{0.8cm} P{0.8cm} P{0.8cm}}
  \hline
  \hline
phase  &  $a$  &  $b$  &  $c$  & $\alpha$  &  $\beta$  & $\gamma$  & $\Delta E$ \\
  \hline
$Fm\overline{3}m$ (c)  & 5.04 & 5.04 & 5.04 & 90.0 & 90.0 & 90.0 & 279.1  \\    
$P$4${\rm{_2}}/nmc$ (t)  & 5.04 & 5.04 & 5.20 & 90.0 & 90.0 & 90.0 & 166.5 \\
$Pca2_1$ (o\rom{3})  & 5.05 & 5.01 & 5.24 & 90.0 & 90.0 & 90.0 &  90.5 \\
$P2_1/c$ (m)  & 5.10 & 5.15 & 5.29 & 90.0 & 80.3 & 90.0 &  0 \\
$Pnm2_1$ (o\rom{4})  & 5.08 & 5.08 & 5.16 & 90.0 & 90.0 & 84.6 & 151.9 \\
  \hline
  \hline
 \end{tabular}
\end{center}
 \caption{Computed lattice parameters and relative energies ($\Delta E$) of different HfO$_2$ phases in their pseudocubic structures, corresponding to the conventional fcc cell of the $Fm\overline{3}m$ structure.
 The lattice parameters given by our DFT calculations match the experimental results very well (see SM section \rom{2}).
 Details of the cell transformations between the primitive cells and pseudocubic cells are provided in the SM section \rom{8}.
 All length, angle, and energy units are in \AA, degrees, and meV/f.u., respectively. 
}\label{t1}
\end{table}

Doping is a well-known technique for stabilizing the t--phase in bulk~\cite{Lee08p012102}.
In thin films, the large surface energy of the m--phase makes it less favorable~\cite{Batra16p172902,Materlik15p134109,Park17p9973}.
As the thickness of the film decreases, the t--phase to m--phase transition temperature is suppressed, and in thin-enough films, the structure remains tetragonal at room temperature~\cite{Shimizu15p032910,Cho12p3534,Kim04p643}.
Recent experimental work demonstrates that the o\rom{3} ferroelectric phase forms as a compromise state resulting from the competition among bulk energy, surface, and doping energies~\cite{Shimizu15p032910,Hoffmann15p072006,Hyuk13p242905}.
Since the t-- and o\rom{3}-- phases have almost equal in--plane lattice parameters (Table~\ref{t1}),  
careful selection of film thickness, doping concentration and growth conditions is needed to stabilize this phase.
For example, it has been proposed that depositing a top electrode before annealing can impose a mechanical confinement and promote the o\rom{3}--phase formation~\cite{Boscke11p102903,Boscke11p112904}.  


Next, we consider the o\rom{4}--phase, which was first proposed theoretically and later investigated in more depth~\cite{Huan14p064111,Sang15p162905}, 
but has not yet been experimentally observed. The o\rom{4}--phase is also ferroelectric, being related to the tetragonal structure by the same polar distortion as the o\rom{3}--phase except along the (110) direction of the primitive tetragonal cell rather than the (100) direction. Our first--principles calculations based on the Berry's phase method~\cite{King93p1651} reveal that this phase has a 0.59 C/m$^2$ polarization along the [011] direction.
We calculated the phonon dispersion for this structure,
finding that there are no unstable phonon modes, which confirms the dynamical stability of this structure (see SM section \rom{3}). Moreover, this structure is elastically stable (see SM section \rom{4}).  

Our DFT results in Table~\ref{t1} show that the relaxed bulk energy of the o\rom{4}--phase is relatively high compared with those of the observed low--temperature (o\rom{3}-- and m--) phases. Furthermore, compared with the t--phase, which is the parent phase of the observed low--temperature phases, 
this ferroelectric phase has a mismatch of about 1\%\ in the in--plane ($a$ and $b$) lattices, and a difference of $5.4^{\circ}$ in the $\gamma$ angle,
making the t--phase to o\rom{4}--phase transition even more unfavorable for a square lattice epitaxial constraint in the (001) orientation.

However, the situation improves for the o\rom{4}--phase in (111)--oriented films. In table~\ref{t2}, we list the lattice parameters of (111)--oriented crystals of different HfO$_2$ phases.
The energies of the $R$3 and $R$3$m$ phases mentioned in Ref.~\cite{Wei18p1095} are also listed for comparison.
Their energies are much higher than that of the high--temperature t--phase, indicating that they are unlikely to form after annealing~\cite{Zhang20p1}.
Both the o\rom{3}--phase and the o\rom{4}--phase have lower energies than the t--phase, with the o\rom{4}--phase having lattice parameters much closer to those of the t--phase; in particular, values of the $a$ and $b$ lattice parameters are the same. The rather small difference in the $\gamma$ angles of the t--phase and the o\rom{4} phase opens the question of the effect of modulating $\gamma$ with an epitaxial shear strain on the relative energy and stabilities of the two phases.

\begin{table}[hpbt]
\begin{center}
 \begin{tabular}{ P{2.0cm} P{0.8cm} P{0.8cm} P{0.8cm} P{0.8cm} P{0.8cm} P{0.8cm} P{0.8cm} }
  \hline
  \hline
phase  &  $a$  &  $b$  &  $c$  & $\alpha$  &  $\beta$  & $\gamma$ & $\Delta E$  \\
  \hline
$Fm\overline{3}m$ (c)  & 7.12 & 7.12 & 8.72 & 90.0 & 90.0 & 120.0 &  279.1  \\    
$P$4${\rm{_2}}/nmc$ (t)   & 7.24 & 7.24 & 8.82 & 88.5 & 91.5 & 121.1 &  166.5 \\
$Pca2_1$ (o\rom{3})   & 7.11 & 7.27 & 8.83 & 91.8 & 90.3 & 119.5 &   90.5 \\
$P2_1/c$ (m)  & 7.25 & 7.38 & 9.47 & 94.9 & 86.6 & 125.4 &   0 \\
$Pnm2_1$ (o\rom{4})  & 7.24 & 7.24 & 9.11 & 91.4 & 88.6 & 123.6 &  151.9 \\
$R$3 & 7.11 & 7.11 & 9.02 & 90.0 & 90.0 & 120.0 & 221.5 \\
$R$3$m$ & 7.13 & 7.13 & 8.72 & 90.0 & 90.0 & 120.0 & 209.6  \\
  \hline
  \hline
 \end{tabular}
  \caption{Computed lattice parameters of different HfO$_2$ phases in their (111)--oriented structures.  We note that there are similar structures whose epitaxial planes are (-111), (-11-1) and so on; for simplicity, these are not included in the list. Detailed lattice parameters and atomic positions are given in the SM section \rom{8}.
 All length and angles units are in \AA~and degrees, respectively.
 } \label{t2}
\end{center}
\end{table}

To explore the shear--induced structural transformation, we start with the distorted t--phase with $\alpha=90^{\circ}$ $\beta=90^{\circ}$, and $\gamma=120^{\circ}$ in its (111)--oriented structure, since the t--phase is the parent phase of the low temperature phases.
The $a$ and $b$ lattice parameters are fixed at the optimized values for the t--phase ($a = b$ = 7.24 \AA), which are the same as those in the o\rom{4} phase.
Next, the $\gamma$ angle is increased from $120^{\circ}$ to $125^{\circ}$ in steps of $0.1^{\circ}$. At each $\gamma$ angle, we relax the $c$ lattice and internal structural parameters while holding $a=b$ and $\alpha=\beta=90^{\circ}$ fixed.  
The energy and out--of--plane polarization are plotted as functions of $\gamma$ in Fig. 2(a).
We can see that as $\gamma$  increases, the local minimum corresponding to the distorted t--structure disappears at  $\gamma=123.5^{\circ}$ and the structure relaxes to the polar o\rom{4}--phase. 
The polar o\rom{4}--phase with $\gamma=123.6^{\circ}$ is shown in Fig. 2(c), and is noticeably different from the (111)--oriented t--phase [Fig. 2(b)].
For $\gamma>123.6^{\circ}$ and up to $125^{\circ}$, the system relaxes to the polar o\rom{4}--structure. 
For this phase, as $\gamma$ decreases, the local minimum corresponding to the polar o\rom{4}--structure disappears at $\gamma=120.7^{\circ}$ and the structure relaxes back to the t--structure.
These results indicate that even though the (111)--oriented t--phase is metastable, 
an in--plane shear skewing its $\gamma$ about $3.0^{\circ}$ can make this non--polar phase unstable to transformation to the o\rom{4}--phase.  

 \begin{figure}[hpbt]
\includegraphics[width=8.5cm]{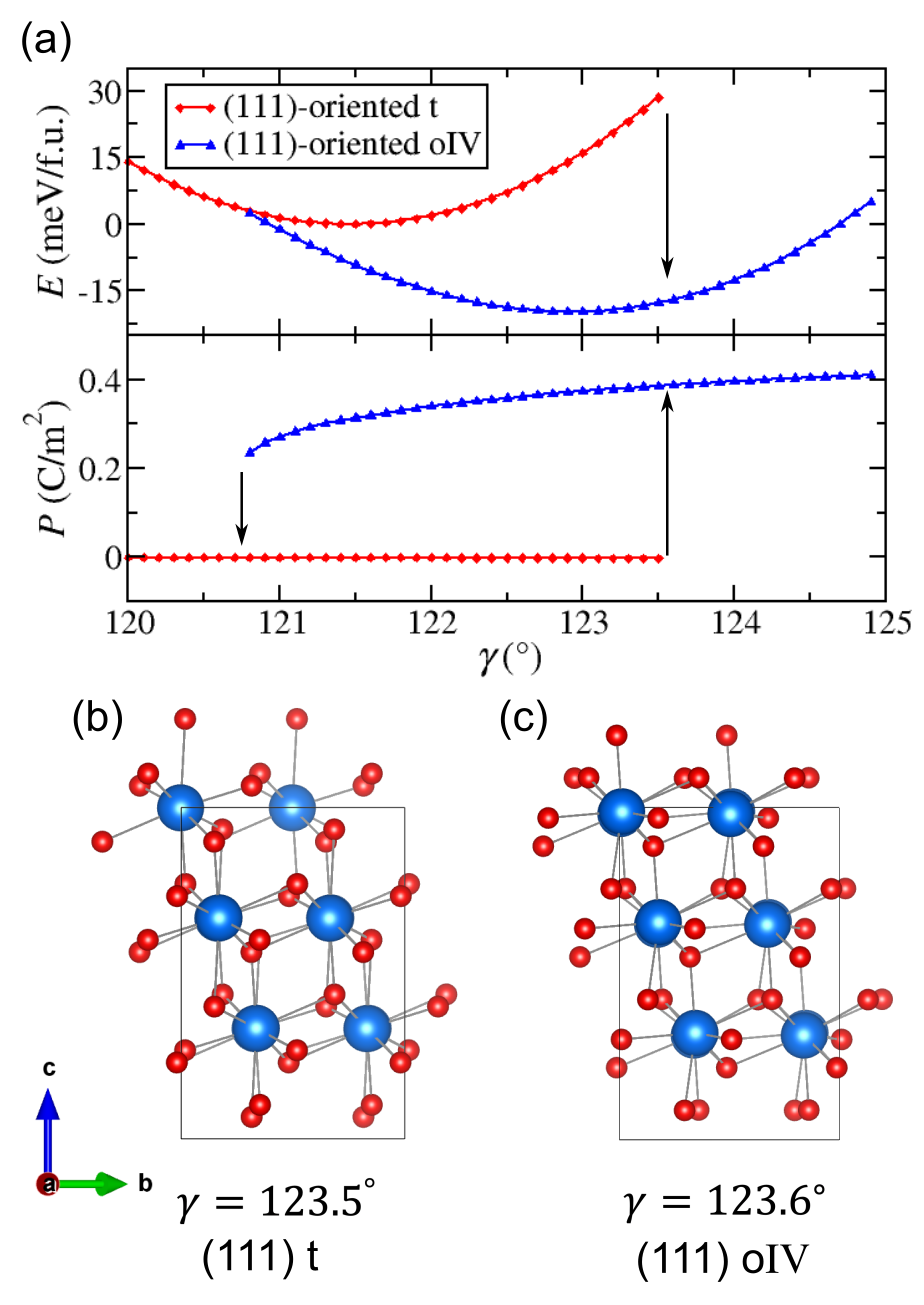}
\caption{(a) Energy and out--of--plane polarization as a function of $\gamma$.
The energy of the t--phase is chosen as the zero of energy. 
(b) and (c) The structures in the t and o\rom{4} phases for $\gamma=123.5^{\circ}$ and $123.6^{\circ}$ respectively.
}
\end{figure}\label{f2}

Based on these results, we propose the following ferroelectric phase
stabilization process. 
At first, a (111)--oriented thin film is deposited on the substrate at a high temperature ($>1000$ K).
At this temperature, a cubic or tetragonal structure is more favorable.
If the substrate is such that the matching condition favors a value of $\gamma$ above the critical value, then on cooling, the t--phase loses its stability, and collapses into the o\rom{4} ferroelectric phase.
It is worth mentioning that under a shear, the m-- or o\rom{3}--phase has a lower energy than the o\rom{4} phase. However, the formation of the o\rom{4} phase is in concert with the Ostwald’s step rule~\cite{Ostwald97p289}, which states that a phase transition may end with a metastable phase with free energy close to that of the parent phase, rather than the most stable phase. In another words, to `kinetically' stabilize a HfO$_2$ functional phase,
it is not necessary to make its energy the lowest. Instead, we can manipulate the reaction path, and promote the phase transition from the high temperature c or t phase to the target phase~\cite{Park19p1800522,Liu19p054404}.

Next, we consider the wake--up effect, which is the need for electric field cycling to establish the full value of switching polarization in as--grown thin films~\cite{Pevsic16p4601,Olsen12p082905}. This wake--up effect has a variety of origins, including internal bias fields~\cite{Schenk15p20224,Schenk14p19744,Mart19p2612}, migration of oxygen vacancies~\cite{Hoffmann15p072006,Starschich16p032903}, structural change at the interface~\cite{Hoffmann15p072006,Grimley16p1600173,Pevsic16p4601}. An important intrinsic factor that can lead to a wake--up effect is an electric field induced non--polar to polar phase transition~\cite{Pevsic16p4601,Hoffmann15p154,Park15p192907,Batra17p4139}.
In the [001]--oriented HfO$_2$--based films, 
the t--, o\rom{3}--, and m--phases are all metastable,
indicating that all three states (or any two of them) can co--exist in an annealed film~\cite{Niinisto10p245,Hyuk13p242905}. 
In thin films with mixed t and o\rom{3} structures, electric field cycling can transform the non--polar t--phase into the ferroelectric o\rom{3} phase, producing a wake--up effect.
However, in the (111)--oriented films, if the axis skew angle is larger than $3^{\circ}$, there is no admixture of nonpolar t--phase as it is unstable to the ferroelectric o\rom{4} phase, which would reduce or eliminate the wake--up effect. 

These results suggest an explanation for the observations in Hf$_{0.5}$Zr$_{0.5}$O$_2$ (HZO) (111)--oriented films ~\cite{Wei18p1095}.
Previous DFT calculations show that HZO and HfO$_2$ have quite similar structural parameters and physical properties~\cite{Materlik15p134109,note1}.
Therefore, it is reasonable to extend our discussion about HfO$_2$ to HZO.
In experiments, HZO was deposited on a STO (001)--oriented substrate at $T=800^{\circ}$C, which favors the non--polar t--phase.
After annealing, thin films with robust polarization (0.34 C/m$^2$ for 5 nm and 0.18 C/m$^2$ for 9 nm) were obtained. 
This ferroelectricity can be attributed to the o\rom{4} phase.
The STO substrate has a square lattice with $C_4$ symmetry, 
 so a coherent epitaxial condition would be $a=b=6.17$ \AA~and $\gamma=127^\circ$ (see SM section \rom{5}).
 While the large difference in lattice parameters means that the film is not expected to be fully coherent, this condition can be understood as a shear favoring larger values of $\gamma$ that destabilize the tetragonal phase as the temperature is lowered.
In our DFT calculation, the out--of plane polarization of (111)--oriented o\rom{4} phase is 0.41 C/m$^2$.
It is not surprising that this value is somewhat larger than the experimentally reported result~\cite{Wei18p1095}, since we consider a perfect polar o\rom{4}--phase in our calculations while the actual thin--films in experiments may contain some or a mixture of the nonpolar crystallites in a part, such as the m--phase. The size of the nonpolar crystallites grows with increasing film thickness, reducing the net polarization in the thicker films. 

 \begin{figure}[h]
\includegraphics[width=8.0cm]{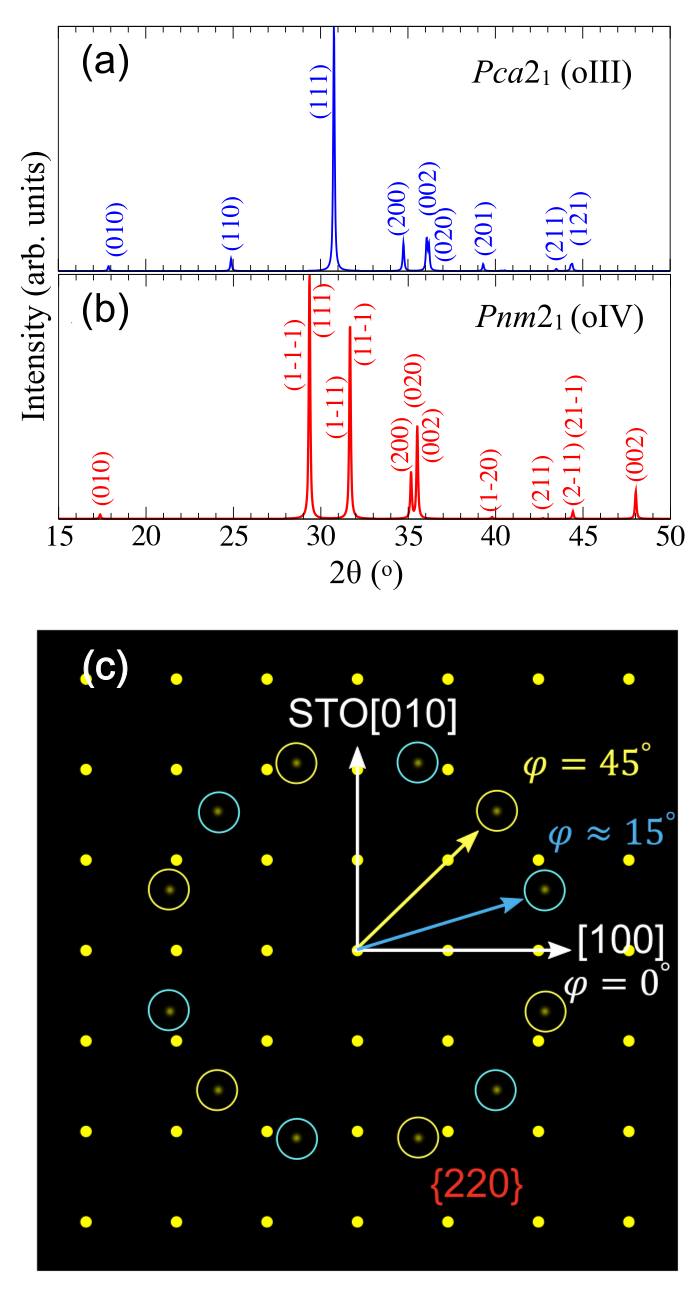}
\caption{Simulated XRD patterns of the bulk (a) o\rom{3} and (b) o\rom{4} structures of HfO$_2$ (see SM section \rom{6} for the comparison with other competing phases, and the XRD patterns in thin films). 
(c) Simulated SAED pattern of a 8.4 nm HfO$_2$ thin film in the (111)--oriented o\rom{4} phase on a STO (001)--oriented substrate. Please refer to SM section \rom{7} for the details about the simulation.}
\end{figure}\label{xrd}

In Fig. 3 (a) and (b), we plot the simulated powder XRD patterns of the o\rom{3} and o\rom{4} HfO$_2$ structures.
First, the (111) peak of the commonly reported o\rom{3} phase lies at $2\uptheta=30.5^\circ$, which is fully consistent with previous experimental reports~\cite{Lyu19p220,Estandia19p1449,Wei18p1095}.
More importantly, the (111)/(1-1-1) peak [$2\uptheta=29.5^\circ$, Fig. 3 (b)] of the o\rom{4} structure lies to the left of the (111) peak of the o\rom{3} phase, which is 
exactly the character of the ferroelectric phase reported in Ref.~\cite{Wei18p1095}.
The secondary peak [(1-11)/(11-1), $2\uptheta=31.5^\circ$] corresponds to the Bragg planes whose normal directions are not out--of--plane. 
Therefore, it is not surprising that such a peak is unobservable in $\theta$--$2\theta$ x--ray measurements~\cite{Widjonarko16p54}.
Here, we would also like to mention that there are several clear features demonstrating the differences between the o\rom{4}-- and m--phase (see SM section \rom{6}).

Fig. 3 (c) shows the simulated SAED pattern of a 8.4 nm HfO$_2$ thin film in the (111)--oriented o\rom{4} phase on a STO (001)--oriented substrate.
Consistent with the experimental observation (Fig. 3 (a) of Ref.~\cite{Wei18p1095}), the simulated SAED also shows that the \{220\} spot has the highest intensity.
Considering symmetry and domains with different lattice orientation, the \{220\} spot leads to a total of 12 different spots, with an approximately $30^\circ$ interval.
The inter--plane distance given by the positions of the spots is $d_{\{220\}}=1.805$ \AA, which is also a remarkable match with the experimental results ($d_{\{220\}}=1.79$ \AA).
We also note that there is a small splitting in each spot in the experimental SAED pattern, which can be explained by a small misalignment of the lattice axes in different domains. 
HfO$_2$ based thin films can adopt various polymorphs with quite similar structures in different circumstances~\cite{Ruh68p23,Wei18p1095,Begon20p043401,Boscke11p102903}.
Here, our analysis shows that many experimental conditions in Ref.~\cite{Wei18p1095},
such as the shear imposed by the substrate, lack of wake--up cycling, and the XRD and SAED patterns observed, match well with our theoretical simulations. 
Therefore, it is reasonable to attribute the ferroelectricity in (111)--oriented HZO thin films on the STO substrate to the formation of the o\rom{4} phase.


In summary, we predict a non--polar to polar phase transition in (111)--oriented HfO$_2$ from first--principles calculations.
Under an in--plane shear strain which enlarges the $\gamma$ angle, the nonpolar $P$4${\rm{_2}}/nmc$ t--phase becomes unstable and relaxes to a polar phase which is a distortion of the ferroelectric $Pnm2_1$ o\rom{4}--phase.
This ferroelectric phase is metastable and has a robust polarization.
We also propose that by a proper selection of the substrate, the wake--up effect, which is undesirable for technological applications, can be reduced or eliminated, so that as--grown films would be ferroelectric.
Our work also offers an understanding the recent experimentally reported ferroelectricity in (111)--oriented Hf$_{0.5}$Zr$_{0.5}$O$_2$ thin films.

{\it Acknowledgements:-- } 
Y.Q., S.S. and K.M.R. were supported by the Office of Naval Research Grant N00014-17-1-2770. 
C.L., F.J.W and C.H.A were supported by the Office of Naval Research Grant N000144-19-1-2104.
F.T.H, X.X. and S.W.C were supported by the center for Quantum Materials Synthesis (cQMS), funded by the Gordon and Betty Moore Foundation’s EPiQS initiative through grant GBMF6402, and by Rutgers University.
Computations were performed by using the resources provided by the High--Performance Computing Modernization Office of the Department of Defense and the Rutgers University Parallel Computing (RUPC) clusters.

\bibliography{cite2}

\begin{thebibliography}{73}
\bibitem [{not()}]{note1}%
  \BibitemOpen
  \href@noop {} {\bibinfo  {journal} {We also carry out DFT calculations for
  HZO with various $\gamma$ angles, and find that HZO shows quite a similar
  shear--induced t to o\rom{4} phase transition.}\ }\BibitemShut {NoStop}%
\end{thebibliography}

\end{document}